# Quantum Emitters in Flux Grown hBN

Evan Williams[1], Angus Gale[1] *, Jake Horder[1], Dominic Scognamiglio[1], Milos Toth[1,2] and Igor Aharonovich[1,2]

[1] *School of Mathematical and Physical Sciences, University of Technology Sydney, Ultimo, New South Wales 2007, Australia*
[2] *ARC Centre of Excellence for Transformative Meta-Optical Systems, University of Technology Sydney, Ultimo, New South Wales 2007, Australia*

* angus.gale@uts.edu.au

## Abstract

Hexagonal boron nitride (hBN) is an emerging material for use in quantum technologies, hosting bright and stable single photon emitters (SPEs). The B-center is one promising SPE in hBN, due to the near-deterministic creation methods and regular emission wavelength. However, incorporation of B-centers in high-quality crystals remains challenging, typically relying on additional post-growth methods to increase creation efficiency. Here, we have demonstrated controlled carbon doping of hBN during growth, using a metal flux based method to increase the efficiency of B-center creation. Importantly, single B-centers with $g^{(2)}(0) < 0.5$ were able to be generated in the as-grown hBN when carbon additions during growth exceeded 2.5 wt.% C. Resonant excitation measurements revealed linewidths of 3.5 GHz with only moderate spectral diffusion present, demonstrating the applicability of the as-grown hBN as a host for high quality B-centers.

## Introduction

Hexagonal Boron Nitride (hBN) has emerged as a promising and desirable material in the field of 2D nanophotonics. The applications are broad, with hBN finding uses in photonic structures and devices [1, 2], phonon-polaritonics[3, 4], as an encapsulation material[5, 6] and spacing layers for heterostructures[7]. Especially important for quantum technologies, hBN also hosts a number of single photon emitters from the ultraviolet (UV) to near infrared (IR)[8-11].

There are several established methods to grow hBN crystals, each with its advantages and drawbacks. Bottom-up fabrication routes include physical deposition molecular beam epitaxy (MBE) [12-15] and chemical vapour deposition (CVD) techniques such as low-pressure (LPCVD)[16, 17] and metal-organic vapour phase epitaxy (MOVPE) [18, 19]. MBE uses sublimated precursor materials to slowly form thin (down to monolayer), high quality and highly-crystalline films at a large scale[20]. However, this comes at the cost of more expensive and non-trivial equipment setups for ultra-high-vacuum work and precursor purities. CVD processes are scalable and flexible, and do not require high-vacuum equipment, allowing them to be freestanding and uncontained. In these methods, a precursor compound is decomposed and thin films formed as the products react on the substrate surface. The most effective growths occur on catalytic metal substrates such as Cu, Fe, Ni and Pt[20,

21]. CVD hBN films are generally of lower quality and polycrystalline, however, with careful control, monolayers and single-crystal films can be grown[22-24]. The requirement of transferring films to dielectric substrates also opens up potential for both physical damage and chemical contamination[25, 26]. MOVPE precursors can be used to overcome this, allowing for higher quality hBN growth on dielectric substrates[27, 28].

The highest quality hBN is produced from bulk crystal growth methods. High pressure high temperature (HPHT) methods utilising GPa pressures and Ba-BN[29] or Mg-BN[30] precursors typically result in the highest quality materials[31]. The main drawback of the HPHT methods is difficulty in scaling. These techniques require large, powerful setups capable of very high pressures and are most suitable for producing crystals on the order of a few millimeters. The atmospheric pressure high temperature (APHT) method is one alternative growth procedure that simplifies equipment requirements and allows for the possibility of increased production scales. The method uses high temperatures at atmospheric pressure to dissolve boron and nitrogen atoms in a chosen metal flux e.g. Fe[32, 33], Ni-Mo[34], Ni-Cr[35, 36] and Fe-Cr[37]. As the flux is slowly cooled, excess boron and nitrogen precipitate from solution and preferentially form hBN. Compared to HPHT methods, experimental setups are simple and crystal sizes can reach centimeter scales with similarly high quality. Pristine quality hBN is usually preferable, but the introduction of carbon as a dopant can be used to increase the density of SPEs[24]. HPHT hBN can be annealed post-growth to introduce carbon at high temperatures[38], while APHT methods have been used to incorporate carbon during the crystal growth [33, 39].

In this work we demonstrate controlled carbon doping of hBN using the APHT flux method. Doping was achieved via the addition of carbon to a single component Fe flux. The methodology in this study is robust, with the same growth conditions utilised for all growths and minimal effect on the quality of the grown hBN for varying carbon levels. The as-grown hBN is also shown to be an excellent host for B-centers, which could be generated in all exfoliated flakes from hBN doped with 2.5 and 5 wt.% C with no additional processing steps required. Resonant excitation measurements were performed at 5 K and linewidths as low as 3.5 GHz were measured, demonstrating the applicability of the as-grown hBN as a host for high quality quantum emitters.

## Results and Discussion

During this study the same single zone tube furnace and experimental conditions were used for all growths, with only the concentration of carbon powder varied. This ranged from 0-5 wt.% C. For clarity, the samples will be denoted by % C only, referring to the wt.% of added C for the rest of the manuscript.
A custom BN crucible was filled with iron and carbon powder and loaded into an alumina tube furnace as seen in Figure 1a. Before any heating took place, the setup was purged for 30 mins with $N_2$ and 5 % $H_2$ in balance with argon to ensure a minimum of contaminants in the system. For the duration of the growth the flow rates were 140 sccm and 25 sccm of $N_2$ and Ar/$H_2$ respectively. The outlet flowed to an exhaust and the system remained at atmospheric pressure for the entire growth. The growth of hBN crystals can be broken down into phases i-iv as follows (Fig. 1b). i) The first phase consists of heating to 1550 °C, at a rate of 5 °C/min. This phase raises the system above the melting point of the iron/carbon mixture resulting in a molten flux. ii) The molten flux is held for 24 hours to dissolve boron and nitrogen atoms from the BN crucible and gas environment. iii) A slow cooling phase then begins at a rate of 2 °C/hr down to 1450 °C. In this phase, as the flux cools, boron and nitrogen precipitate out and preferentially bond in a hexagonal configuration. With the addition of carbon in the melt, as the crystal forms, carbon atoms are incorporated into the hBN lattice. A rate of 2 °C/hr was chosen for this study to ensure sufficiently large crystal domains. Faster rates, up to 10 °C/hr, have been demonstrated in literature to decrease the overall growth time, however this comes at the cost of smaller crystal domains[36]. At the completion of this phase, the temperature of 1450 °C ensures that the flux is solidified and growth of hBN is complete. iv) The final phase is a rapid cool to RT at a rate of 5 °C/min, at which point the steel ingot with hBN can be recovered (Fig. 1c). The conditions chosen above allow for the same procedure to be used for both pure hBN (0 % C) and carbon inclusions up to 5 %.

The lateral size of the carbon doped hBN in this study was limited by the crucible. The resultant ingots are on the order of 20 - 30 mm in length. Given a larger furnace and crucible, the method could be easily scaled. To isolate the grown hBN crystals, mechanical cleavage is used to remove them from the solid metal ingot. Exfoliation of hBN flakes is then straightforward, using tape-based methods onto a chosen substrate as required. In Figure 1d, hBN flakes were exfoliated onto an Si/SiO$_2$ substrate for further analysis. The flakes display a range of thicknesses, apparent from the colour, and lateral sizes up to ~200 µm, on the order of the size of the crystal domain.

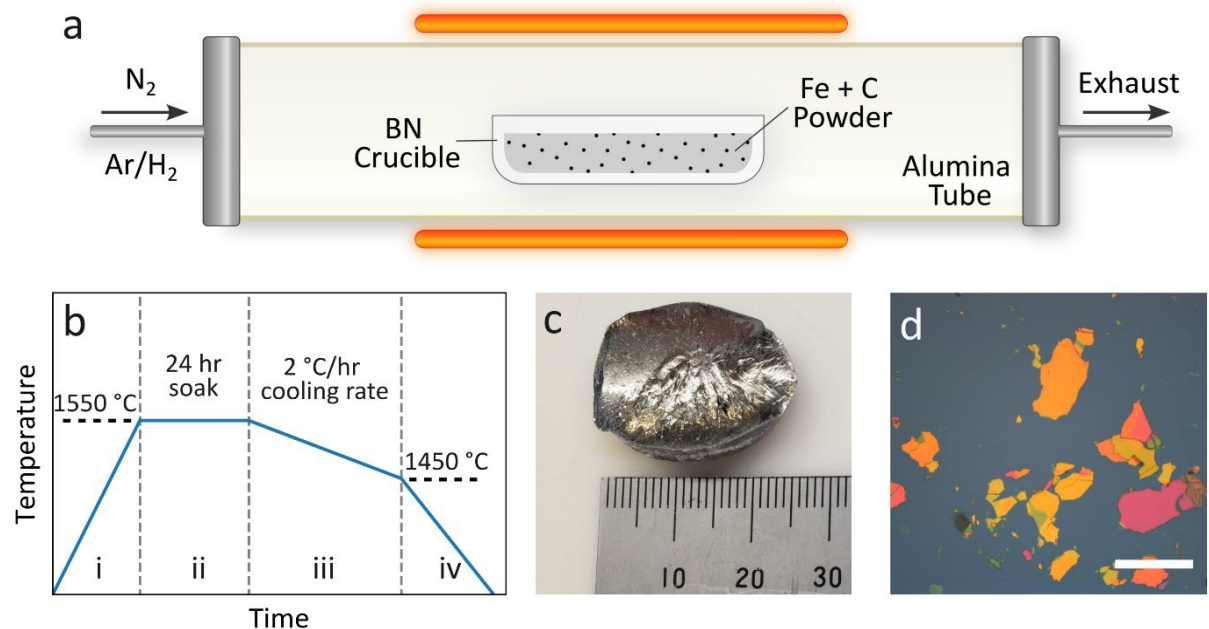

*Figure 1: Overview of the experimental setup to grow carbon doped hBN. a) Diagram of growth setup and crucible within the single zone tube furnace – The carbon inclusions are scaled as required from 0 - 5 % C. b) Plot of the temperature profile utilised for all growths. Details including the phases (i - iv) are included as necessary but note the plot is not to scale. c) Metal ingot removed from BN crucible post-growth using a 5 % carbon inclusion. Scale units are millimeters. d) Optical image of hBN flakes transferred to SiO$_2$ substrate via direct scotch tape exfoliation method. 5 % C Crystals were exfoliated from the ingot shown in c). The scale bar represents 200 µm.*

In the ideal case, any carbon inclusions in the hBN will be incorporated into the lattice of the hBN with a minimal disruption of the crystallinity of the material. Optical images show the crystal domains of each hBN sample before removal from the metal ingot (Fig. 2a). Raman spectroscopy was employed to characterise the as-grown hBN samples. In all grown samples the full width at half-maximum (FWHM) is ~8 cm$^{-1}$ (Fig. 1b). This is indicative of a highly crystalline material, on par with other high quality hBN sources in literature[31]. The narrow FWHM indicates that even at the 5 % C inclusion in this work, the resultant hBN remains highly crystalline.

Cathodoluminescence (CL) measurements were also utilised and serve several purposes in the context of this work. CL can probe for defect emissions across a broad range of the electromagnetic spectrum. Emissions from the well-studied UV defect, with ZPL at 4.1 eV, are also indicative of the presence of carbon in the hBN crystal[29, 33]. Additionally, the activation of the B-center can be monitored in real-time[40]. Figure 2c shows representative room-temperature CL spectra from exfoliated flakes using a focused electron beam (5 kV, 3.2 nA). For all samples the characteristic band edge emission from hBN is not present due to transmission losses in the collection path of the CL setup. In both the undoped and 1 % C samples, relatively broad emissions centered around 300 nm are present. Band edge emissions in hBN emit at ~210-215 nm, so these lower energy emissions may stem from defects formed during the growth of hBN[31]. A number of emissions in this spectral region, unrelated to the 4.1 eV defect, have been noted in other studies[39, 41]. For the 2.5 and 5 % C samples the typical 4.1 eV emission is clearly visible. The appearance of this emission only at 2.5 % C and above may be related to the ability of the iron solvent to retain carbon below a particular concentration, acting as a "carbon sink"[35]. As there is clear experimental evidence suggesting the 4.1 eV defect is carbon based, emission from this defect is a strong indication that carbon has been incorporated into the hBN crystals[29, 33, 38].

In addition to the 4.1 eV emission there is also the appearance of the B-center emission with ZPL at 436 nm. This emission is created during CL acquisitions as the measured spot is irradiated with focused electrons. The presence of the 4.1 eV emission is a prerequisite for B-center activation and before any electron irradiation these emissions are not present in the as-grown hBN[42]. For both the 2.5 and 5 % C samples, these defects were activated in all exfoliated flakes without requiring any additional processing methods such as annealing. While both the 0 and 1 % C samples had some other UV emissions, the B-center emission was never measured even with long dwell times.

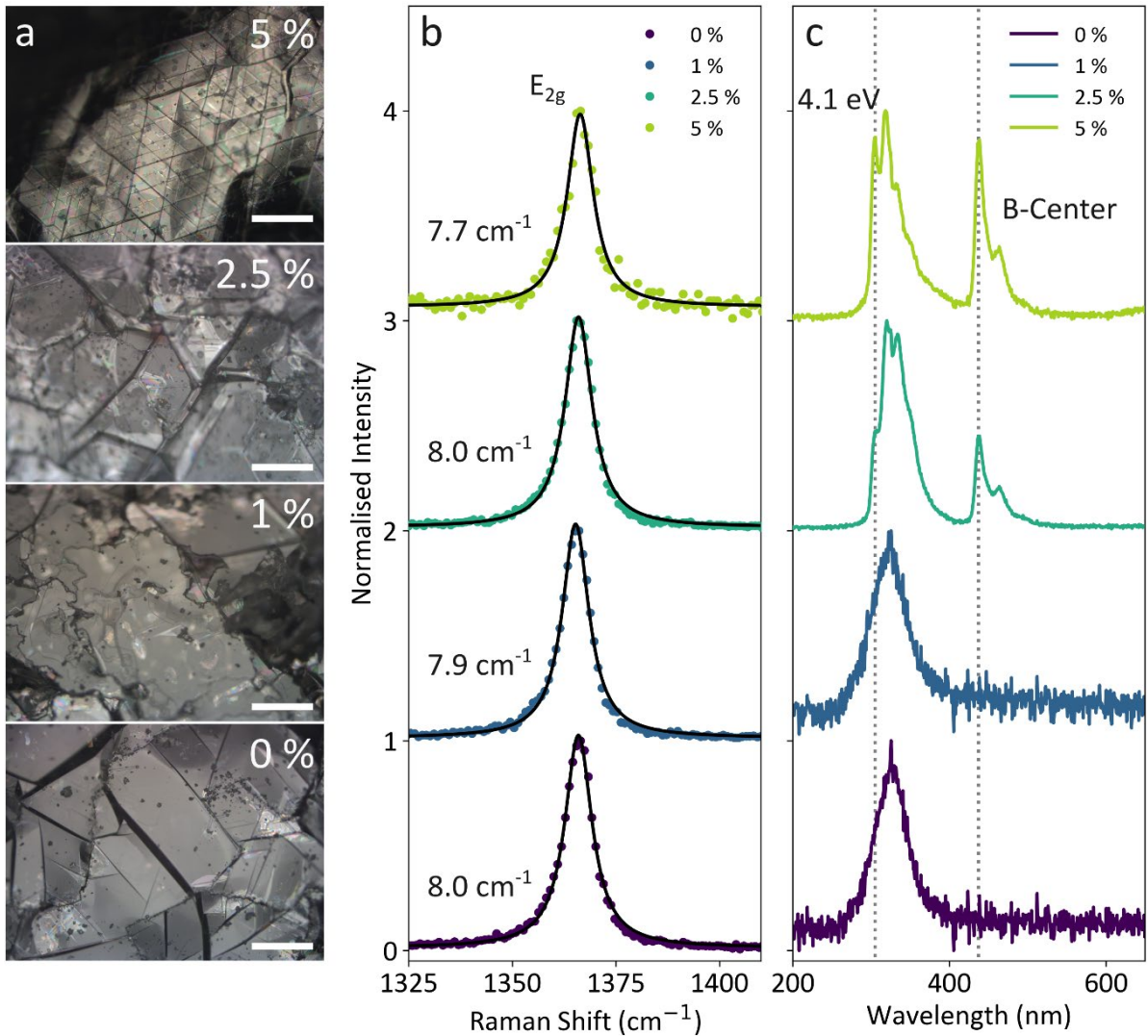

*Figure 2: Characterisation of crystal quality and defect inclusion by impurity concentration. a) Optical images of crystal domains on the metal ingots at each level of C-inclusion – 0 %, 1 %, 2.5 % and 5 %. The scale bars represent 200 μm. b) Raman spectra of the hBN $E_{2g}$ mode at 1366 cm$^{-1}$, measured directly from crystals on the surface of the Fe ingots. The FWHM are labelled for each level of C-inclusion. c) CL spectra of hBN flakes for each level of C-inclusion. The characteristic carbon-related 4.1 eV defect and B-center emissions are labelled with dotted lines indicating the position of the ZPLs.*

The presence of the 4.1 eV emission and the ability to activate B-centers in both the 2.5 and 5 % C hBN samples, as-grown, is promising for quantum applications. Other techniques to enhance the B-center activation efficiency typically use annealing procedures, which add complexity to workflow. In order to further characterise the B-centers hosted within the grown hBN, selected flakes in both the 2.5 and 5 % C samples were irradiated with focused electron irradiation to generate B-centers. An array was patterned using focused electron irradiation in an SEM, with doses ranging from 2.0 x 10$^9$ - 6.4 x 10$^{10}$ electrons/spot. Room temperature photoluminescence (PL) spectra from a single B-center in each flake was measured with the typical ZPL and PSB emissions present (Fig. 3a,b). The cutoff in spectral intensity above 490 nm is due to a band pass filter in the collection. Autocorrelation measurements confirmed each emitter was a single isolated defect with g$^{(2)}$(0) = 0.25 for the 2.5 % C sample (Fig. 3a) and 0.10 for the 5 % C sample (Fig. 3b). Lifetimes extracted from the g$^{(2)}$(τ) plot are typical for the B-centers, with values of 2 ns and 1.6 ns for the 2.5 and 5 % C samples respectively.

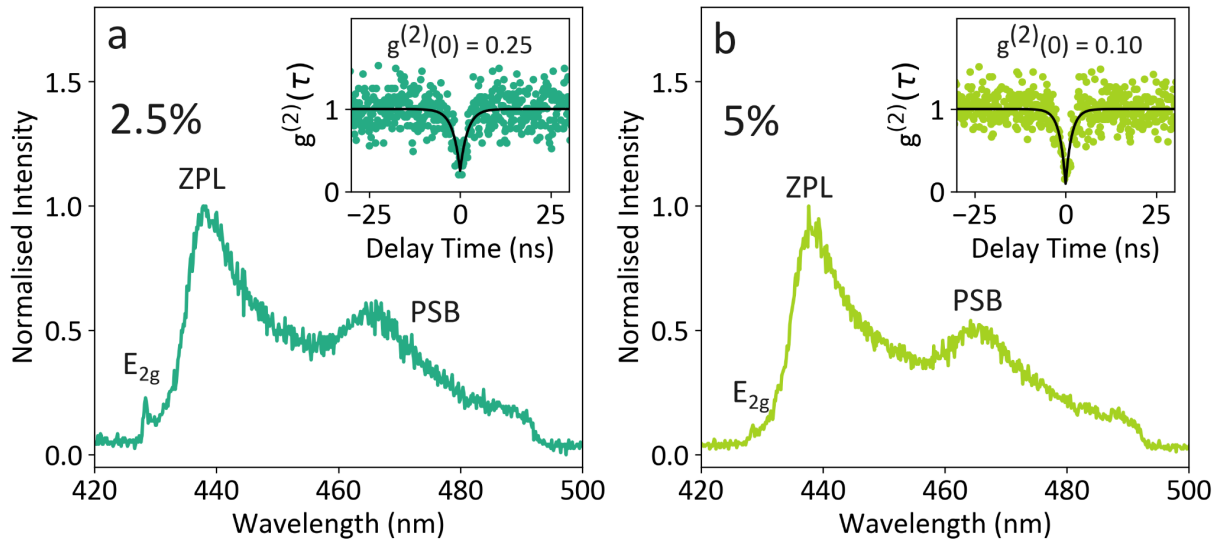

*Figure 3: PL and autocorrelation measurements of emitters in C-doped hBN. a) Room temperature B-center PL spectrum and (inset) g2 autocorrelation function from a 2.5 % C flake. b) Room temperature B-center PL spectrum and (inset) g2 autocorrelation function from a 5 % C flake. The hBN E$_{2g}$ Raman line, ZPL and PSB emissions are labelled on each spectrum.*

For quantum applications, an ideal single photon source should be bright and have near lifetime limited linewidths. Spectral diffusion and phonon broadening can significantly increase the measured linewidths of single emitters in hBN. Linewidth broadening from electron-phonon coupling can be minimised by cooling samples to cryogenic temperatures, typically < 10 K. Resonant excitation schemes can then be used to quantify any effects of spectral diffusion on the measured linewidths, which are particularly pronounced for B-centers in hBN[43]. In this work, B-centers in a 2.5 % C hBN flake were cooled to 5 K, and a tuneable laser was used to resonantly excite the transition between ground and excited states.

As mentioned previously, B-centers are efficiently created in carbon-doped samples. Homogenous linewidths for the B-centers are on the order of 100 MHz[44], but defects and charge traps surrounding the defect increase spectral diffusion significantly, such that the measured linewidths are on the order of 1 GHz, even in the highest quality hBN[45]. The broad linewidths caused by significant spectral diffusion, means that implementation in many applications that require highly coherent photons is not trivial. For hBN samples doped with carbon using post-growth annealing, the average FWHM of a B-center is ~15 GHz[45]. Figure 4a shows the power-dependent ZPL broadening of the emitter from 10 nW - 30 µW. Multiple linescans are recorded and the intensity trace fitted with a Lorentzian function to extract maximum intensity and FWHM. Figure 4b shows the saturation curve of the emitter. The data was fit with a saturation function I(P) = I$_\infty$ P / (P + P$_{sat}$) from which the ultimate intensity I∞ and saturation power P$_{sat}$ are found to be 117 kcps and 8.57 µW respectively.

The B-centers in the 2.5 % C sample shows only moderate spectral diffusion effects with a power independent linewidth Γ$_0$ of 3.52 GHz extrapolated using the fitting function Γ(P) = Γ$_0$ (1 + P/P$_{sat}$)$^{1/2}$ (Fig. 4b). The incorporation of carbon during crystal growth is expected to reduce any high strain or damage subjected to crystals during post-growth techniques, leading to a reduction in spectral diffusion as shown in these as-grown samples.

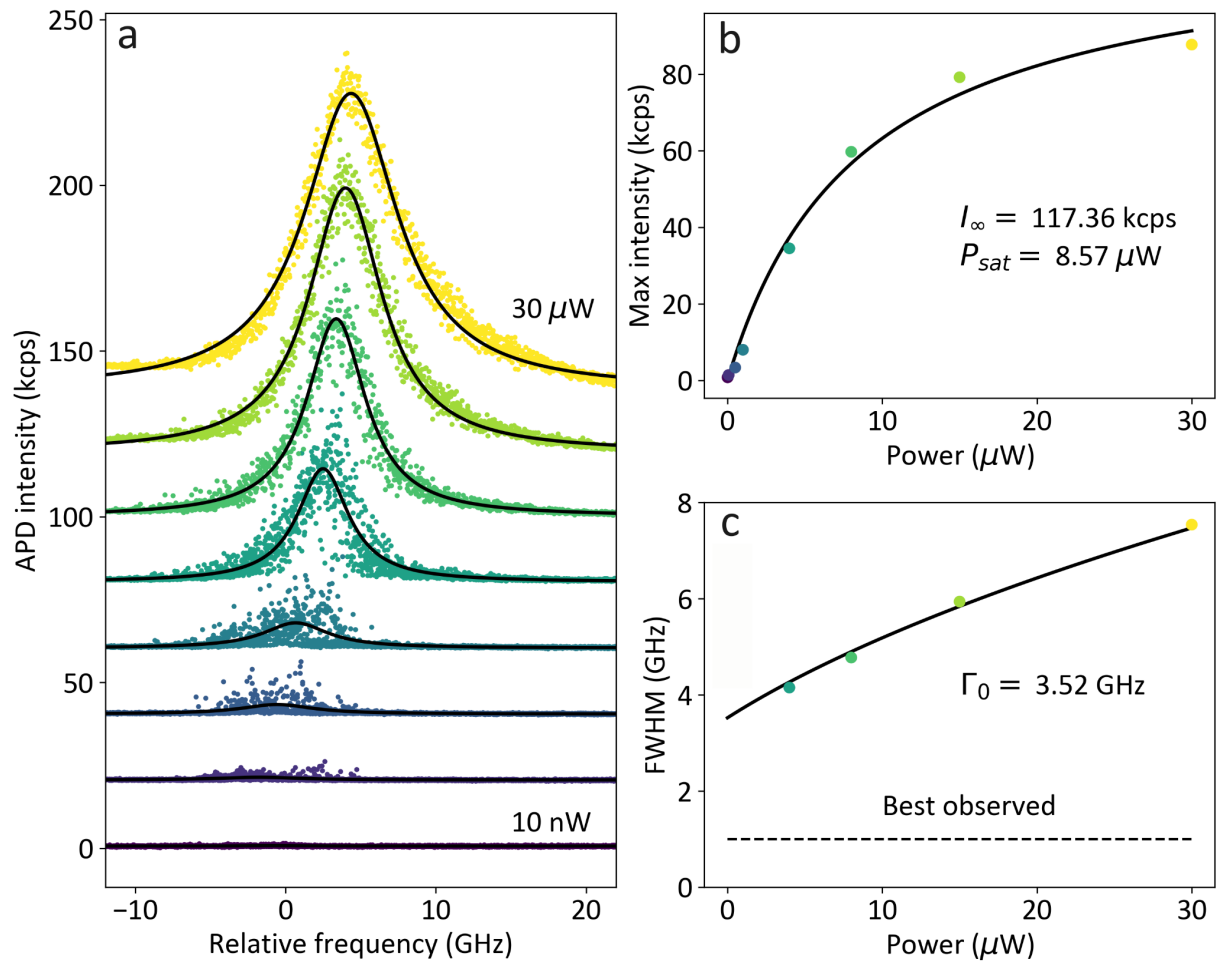

*Figure 4: Characterisation of a representative B-center in a 2.5 % hBN flake under resonant excitation. a) Linescans across the ZPL at increasing laser power, from 10 nW to 30 µW. For each power the data are fit with a Lorentzian function, from which the maximum intensity and linewidth are extracted. b) The maximum intensity from the fits in a) plotted as a function of power. A saturation function was fit to the data. The extracted value for ultimate intensity I∞ is 117 kcps and saturation power P$_{sat}$ is 8.57 µW. c) The linewidth from the fits in a) follows a typical broadening behaviour under increasing power. A fit to the data gives a power independent linewidth of 3.52 GHz. Data for powers below 1 µW were excluded from the plot and fit due to relatively high noise. The dotted line corresponds to the best observed linewidths for B-Centers in hBN, shown for comparison.*

# Conclusion

In this study we have demonstrated the ability to carbon dope hBN using the APHT flux method and control carbon inclusions in the melt. The methodology is robust and the carbon doped crystals are of high quality as demonstrated by Raman spectroscopy. Large crystal domain sizes of the grown hBN allow for easy exfoliation of hBN flakes onto Si/SiO$_2$ substrates. The presence of the UV defect emitting at 4.1 eV indicated that carbon was successfully incorporated in 2.5 and 5 % C samples. Single B-centers were successfully created in exfoliated flakes from these samples and by introducing carbon during the crystal growth we have avoided significant spectral diffusion present in post-treated crystals. The ultimate linewidth of 3.5 GHz for a B-center created in these flakes is significantly narrower when compared to those in post-doped hBN. Our work is promising for incorporating quantum defects into hBN during growth for emerging quantum applications, without sacrificing the crystal quality. In the future, we envision utilisation of a similar method to explore other dopants within the hBN lattice.

# Methods

hBN Crystal Growth:
hBN crystals were grown using the metal solvent method in an alumina tube furnace. Iron powder (>99 % purity) was loaded into a custom hot-pressed boron nitride crucible. For carbon doped samples, graphite powder of the specified wt.% was mixed into the iron before loading. After placing the crucible into the furnace, the system was purged for 30 minutes with a combination of 140 sccm of nitrogen gas and 25 sccm of 5% hydrogen in balance with argon, both controlled using digital mass flow controllers. The tube was then sealed and filled with the same ratio of gas to raise the pressure to atmospheric. Once at pressure the bypass valve was opened, allowing gas to exit the chamber at a passive rate as new gas was pumped in. The furnace was heated to 1550 °C at a rate of 5 °C/min and left there for 24 hours. The furnace was then cooled at a rate of 2 °C/hr to 1450 °C, where it then cooled at 5 °C/min to ambient temperature.

Flake transfer:
hBN flakes were transferred using tape exfoliation. Crystals were mechanically removed from the ingot surface and placed onto 3M scotch tape. Si/SiO$_2$ (285 nm oxide thickness) substrates were sonicated in acetone for 15 minutes, and then Isopropyl alcohol for a further 15 minutes before being dried with nitrogen. Directly prior to exfoliation they were then placed in an ozone plasma cleaner for 30 minutes. After exfoliation, substrates were placed on a hot plate at 500 °C to remove any tape residue.

Raman Characterisation:
Raman characterisation was performed with a Renishaw inVia Raman spectrometer using a Leica DMLB microscope and a 633 nm laser (He-Ne source).

Cathodoluminescence measurements and emitter generation:

CL measurements were taken using a custom-built system using a modified gas injection port on a Thermo Fisher Helios 5 PFIB DualBeam microscope. A focused electron beam was utilised for all measurements with 5 keV accelerating voltage and 3.2 nA beam current. A ball lens was used to focus emissions into a 600 µm optical fibre. Spectra were collected using an Ocean Optics 65000 spectrometer. B-centers were created using the same conditions and spot arrays patterned using the inbuilt patterning software.

Photoluminescence measurements:

Room temperature PL measurements were taken using a custom lab-built confocal microscope setup. Excitation was performed with a 405 nm laser. Collection was filtered using a 460 ± 30 nm band-pass filter and collected using an Andor Kymera 328i spectrometer (for spectra) or 50/50 beam splitter fiber to Excelitas avalanche single photon detectors (for autocorrelation measurements). Cryogenic photoluminescence measurements were taken at 5 K using a closed-loop cryostat. Resonant excitation was performed with a frequency doubled Ti:sapphire scanning laser and photons were collected from the phonon sideband using a 442 nm long pass filter and multimode fiber. Data was collected over 25-ms integration time with the laser wavelength scanned at a rate of 1 GHz/s.

# Acknowledgements

We acknowledge financial support from the Australian Research Council (CE200100010, FT220100053, DP240103127) and acknowledge the UTS node of the ANFF for access to nanofabrication facilities. The authors would like to thank Professor James Edgar for his valuable discussions on crystal growth methodology, Dr Toby Shanley for assistance and maintenance of the experimental setups and Dr Mika Westerhausen for useful discussions on chemistry and concepts.